# Spin waves interference from rising and falling edges of electrical pulses

Jae Hyun Kwon,[1] Sankha Subhra Mukherjee,[1] Mahdi Jamali,[1] Masamitsu Hayashi,[2] and Hyunsoo Yang[1,a]

[1]*Department of Electrical and Computer Engineering and NUSNNI-NanoCore, National University of Singapore, 117576, Singapore*

[2] *National Institute for Materials Science, Tsukuba 305–0047, Japan*

The authors have investigated the effect of the electrical pulse width of input excitations on the generated spin waves in a NiFe strip using pulse inductive time domain measurements. The authors have shown that the spin waves resulting from the rising- and the falling-edges of input excitation pulses interfere either constructively or destructively, and have provided conditions for obtaining spin wave packets with maximum intensity at different bias conditions.

[a] e-mail address: eleyang@nus.edu.sg

Although yttrium iron garnet (YIG) has provided a great vehicle for the study of spin waves in the past, associated difficulties in film deposition and device fabrication using YIG had limited the applicability of spin waves to practical devices. However, microfabrication techniques have made it possible to characterize both the resonant[1] as well as the travelling characteristics[2, 3] of spin waves in permalloy (Py). A variety of methods have been used for measuring spin waves, including Brillouin light scattering (BLS)[4], magneto-optic Kerr effect (MOKE)[5], vector network analyzer ferromagnetic resonance (VNA-FMR)[6], and pulse inductive microwave magnetometry (PIMM)[1, 7].

PIMM is one of the most preferred methodologies of measuring travelling spin waves. In this method, an electrical impulse is applied at one of two coplanar waveguides patterned on top of oxide-insulated Py, producing a local disturbance in the magnetization of the Py. The resulting disturbance travels down the Py in the form of waves, and is inductively picked up by the other coplanar waveguide. This technique lends itself most amenable to digital information transfer and processing using the interference of spin waves.[8] It must be noted that, for practical applications, one needs to use impulses rather than either the rising/falling edge of applied electrical signals, because spin waves resulting from the rising edge of a pulse generally differ in magnitude from that of the falling edge of the pulse due to differences in the rise/fall times of most pulse generators, and also generally result in significant interference in subsequent measurements. In most reports, however, little has been mentioned regarding the characteristics of the finite pulse width of excitation pulses on the generated spin waves. In fact, Covington *et al.* mentioned that there is little difference between spin wave data obtained from impulse and step excitations.[2] Barman *et al.* first reported qualitatively the coherent destruction of spin waves resulting from pulses having certain pulse widths.[9]

In this letter, we investigate the effect of the pulse width of excitation pulses on the generated spin wave packets using both experimental results and micromagnetic simulations. We show that spin wave packets generated from electrical pulses are a superposition of two separate spin wave packets, one generated from the rising edge and the other from the falling edge, which interfere either constructively or destructively with one another, depending upon the magnitude and direction of the field bias conditions. An analytical expression has also been provided, relating the pulse width to the degree of degradation/enhancement of a signal, and may be used for determining the excitation pulse widths which result in maximum resultant spin wave intensity.

Figure 1(a) shows an optical micrograph of the device with a 20 nm thick $Ni_{81}Fe_{19}$ (Py) strip (150 × 30 μm$^2$), deposited on a Si/SiO$_2$ (100 nm) substrate. The Py strip is subsequently covered with 30 nm of SiO$_2$, and finally Ta (5 nm)/Au (85 nm) is sputtered on top and patterned into asymmetric coplanar strips (ACPS). Voltage pulses are applied to the left ACPS by a pulse generator having a rise time of 75 ps and a fall time of 88 ps, and the induced voltage resulting from travelling spin wave packets are detected at the other ACPS by a real-time oscilloscope. For noise reduction, the signal is passed through a 20 dB low noise amplifier and averaged over 10,000 measurements. An out-of-plane bias magnetic field ($H_b$) was applied during the measurements.

Figure 1(b) shows an example of the 2 V applied voltage pulse with a pulse width ($t_δ$) of 10 ns by a green line, and the detected signal by a blue line. The spin wave signal is measured at an $H_b$ of 800 Oe. Each measurement dataset at a fixed $t_δ$ has been obtained by making measurements at -5.9 kOe, 800 Oe, and 5.9 kOe, in that order. Then, the signal due to the wave packet obtained for the measurement at -5.9 kOe is subtracted from that obtained at 800 Oe, so that the background signal is removed. Finally filtering is performed in the frequency domain, so that the high frequency spin wave signal at 5.9 kOe is removed from the final measurement. In Fig. 1(c), a contour plot of the measured time domain signals is shown as a function of $H_b$ and time, for a 2 V, 100 ps pulse input. The FFT of the time domain signal is shown in Fig. 1(d). The variation of the frequency of the measured signal with $H_b$ is a clear indication that the measured signals are due to spin waves.[2,10]

A careful examination of Fig. 1(b) reveals that the spin waves resulting from the rising and the falling edges of the input rectangular wave are phase shifted from one another by $\pi$ radians. This is easily understood if one observes that the first peak of the wave packet resulting from the rising edge of the input is positive, while that resulting from the falling edge is negative. From this observation, it is possible to construct representative equations for the Gaussian wave packets arising due to the rising and falling edges as $\xi_r(t) = A_r\cos(2\pi ft)\exp(-t^2/[2\sigma^2])$ and $\xi_f(t) = A_f\cos(2\pi f[t-t_\delta]+\pi)\exp(-[t-t_\delta]^2/[2\sigma^2])$, respectively, where $A$ is the amplitudes of the wave packets, $f$ is their frequency, $\sigma$ is the standard deviation, and $t_\delta$ is the width of the input pulse. The waves constructively interfere under the condition $2\pi ft_\delta+\pi = 2n\pi$, for integer $n$, whereas they destructively interfere when $2\pi ft_\delta+\pi = (2n+1)\pi$. In fact, the expression $[ft_\delta-Int(ft_\delta)]$, where $Int(x)$ refers to the integral part of the real number $x$, is a measure, in a scale of zero to one, of the degree of destructive interference.

In Fig. 2 (a-c), spin wave packets arising due to the rising and falling edges of a rectangular voltage pulse at a bias field of 800 Oe, with an amplitude of 2 V and zero offset voltage (a unipolar pulse) is shown. Figure 2(a) depicts line plots of measured spin wave packets arising from rectangular input voltage pulses for various values of $t_\delta$. The wave packet at approximately 4 ns corresponds to that generated due to the rising edge of the input voltage pulse. The other wave packet which steadily shifts to the right for line plots staggered higher is due to the falling edge of the pulse. The lines are color-coded from blue to red representing the degree of destructive interference as mentioned above, blue being destructive interference, while red being constructive interference. The interference is better visible in contour plots in Fig. 2(b) and 2(c), where the amplitude of the measured spin wave packets is plotted as a function of $t_\delta$ and the measurement time ($t$). The set of vertical lines around 4 ns again represents the wave packets arising due to the rising edge of the input signal. The set of lines at an angle represents spin wave packets resulting from the falling edge of the input. When zoomed into a section in which $t_\delta$ is less than 2 ns (since the spin wave packets decay within that time) in Fig. 2(c), the interference of the waves are clearly visible. Constructive interference occurs at pulse widths of 0.33 ns, 0.99 ns, and so on, while destructive interference occurs at 0.66 ns, 1.32 ns, and so on. A voltage pulse that has an amplitude of 2 V but also an offset of -1 V is defined as a bipolar pulse. Spin wave packets arising from bipolar pulses are shown in Fig. 2(d-f) respectively and results are similar to the unipolar case. The magnitude and frequencies are slightly different due to the different magnetization directions between unipolar and bipolar excitations. Constructive interference for bipolar pulses occurs at 0.38 ns, 1.16 ns, and so on, while destructive interference occurs at 0.76 ns, 1.92 ns, and so forth.

To better understand the effect of excitation pulse width on the spin wave packet, micromagnetic simulations have been performed using OOMMF.[11] The structure used in the simulations is 6 μm in length, 900 nm in width, and has a thickness of 20 nm. The dimensions were selected so as to have the same aspect ratio of length vs. width as the measured sample. A 2 kOe magnetic field was applied perpendicular to the sample as a bias field. The simulation cell size was $10\times10\times20$ nm$^3$ with the ferromagnetic layer, and made of Py with a saturation magnetization ($M_s$) of $860\times10^3$ A/m, the exchange stiffness ($A_{ex}$) of $1.3\times10^{-11}$ J/m, and a Gilbert damping constant ($\alpha$) of 0.01. In order to generate spin waves, an external magnetic field was applied to a 10×900×20 nm$^3$ area at the center of the ferromagnetic structure. The applied field has a fixed rise time and fall time of 100 ps and the field pulse width was varied from 100 up to 800 ps. As can be seen from the $M_x/M_s$ component in Fig. 3(a), the magnetization stays in the plane of the film at center of the wire. An absorbing boundary condition has been applied to the ferromagnetic wire edges to avoid spin wave reflection.[12,13] The spin wave intensity was calculated by averaging the magnetization in a 200×900×20 nm$^3$ area, 2 µm away from the spin wave source. The time response of the spin wave for different excitation pulse widths is shown in Fig. 3(b). As similar to the experimental result, the interference of the spin wave between the one launching from the rising edge of excitation and the other one from the falling edge could be observed. By increasing the excitation pulse width, these two spin waves interfere less and the interference is almost negligible above 1 ns pulse width.

The concept that has been introduced is general and could be applied to other spin wave modes. We have simulated spin wave packets for the same structure, when the bias field is in-plane and along the nanowire. Figure 3(c) shows the $M_x/M_s$ component of magnetization, when a 1 kOe field is applied along the ferromagnetic nanowire. The output spin wave for different excitation pulse widths is shown in Fig. 3(d). The interference pattern due to the rising and falling edges on the resultant spin waves are clearly visible. However, due to a different spin wave mode, the frequency of the generated spin wave is higher than that of Fig. 3(b), and thus the interference occurs at different pulse widths in comparison to the out-of-plane bias field.

Another point of view of the resultant interference pattern would be the following. The square excitation pulse is composed of many sinusoidal components at different frequencies, which may be obtained by a Fourier transform of the excitation signal in the time domain. The component of the excitation signal responsible for resonant spin dynamics would be the one at the resonant frequency $f$ of the ferromagnet as determined by the bias field conditions. The Fourier transform of a square pulse of pulse width $t_\delta$ is given by $\text{sinc}(ft_\delta)$, where $\text{sinc}(x) = \sin(\pi x)/(\pi x)$. The magnitude of the interfering signals in Fig. 4(a) due to unipolar excitation is compared with the amplitude of the component of the excitation signal at a $f$ of 1.5 GHz in Fig. 4(b), showing good agreement with the experimental data.

In conclusion, the authors have shown that spin waves generated by electrical pulse excitations are composed of the superposition of two spin wave packets, one resulting from the rising edge and the other from the falling edge of the excitation pulse. It is further demonstrated that the spin wave packet resulting from the rising and falling edge are phase-shifted by 180º. For excitation pulse width less than the standard deviation of the resultant spin wave packets, these spin waves interact either constructively or destructively with one another. The phenomenon is shown to be independent of the type of pulse excitation and the mode in which the spin waves are generated.

J.H.K. and S.S.M. contributed equally to this work. This work is supported by the Singapore NRF CRP Award No. NRF-CRP 4-2008-06.

Figure captions

Fig. 1. (a) An optical micrograph of the device. (b) A square wave input produces Gaussian spin wave packets at both the rising and falling edges. (c) Contour plot of measured spin wave packets as a function of the bias field ($H_b$) for a 100 ps pulse excitation. The scale bar is in mV. (d) The change in spin wave frequency as a function of $H_b$.

Fig. 2. (a) Line plots of Gaussian wave packets due to the rising and falling edge of a unipolar voltage pulse for the various pulse widths, $t_\delta$ showing constructive (red lines) or destructive (blue lines) interferences. (b) Contour plot showing the two Gaussian wave packets as a function of $t_\delta$. (c) A section of (b) zoomed in to reveal the interference. (d), (e) and (f) are plots corresponding to (a), (b) and (c) for a bipolar pulse, respectively. The scale bar is in mV.

Fig. 3. The $M_x/M_s$ component (a) and the time response of the spin wave for different $t_\delta$ (b) with a 2 kOe out-of-plane bias field. The $M_x/M_s$ component (c) and the spin wave for different $t_\delta$ (d) with a 1 kOe in-plane bias field along the nanowire.

Fig. 4. (a) The contour plot of spin waves generated from the rising and falling edges of a unipolar square pulse with different $t_\delta$. (b) The amplitude of the frequency spectrum of a square pulse with different $t_\delta$, at the resonance frequency $f$ of 1.5 GHz.

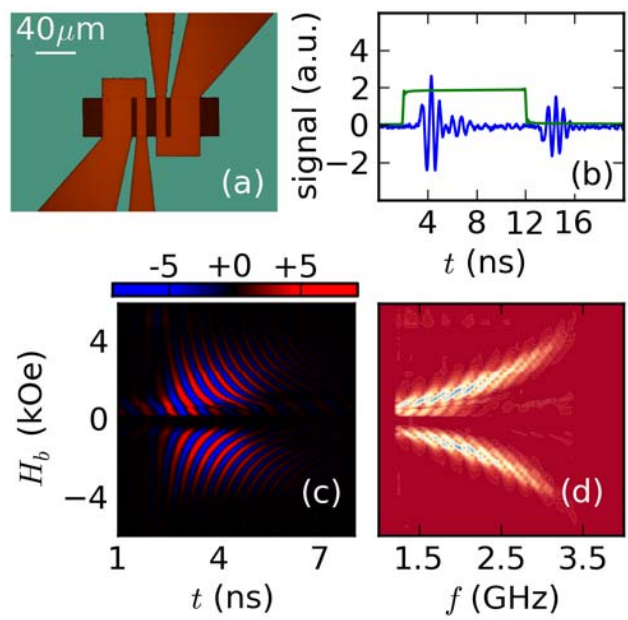

Figure 1.

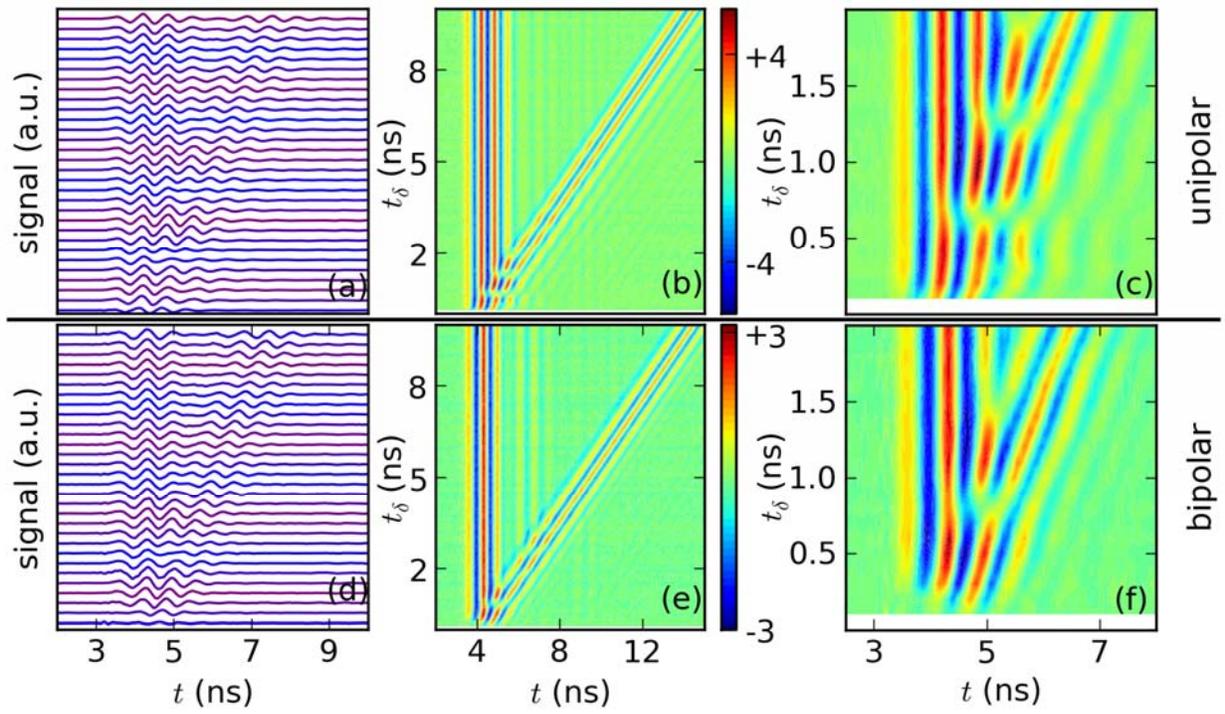

Figure 2

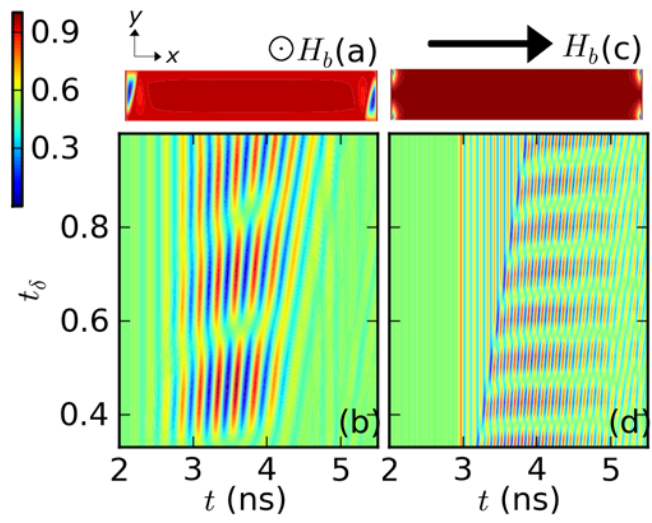

Figure 3.

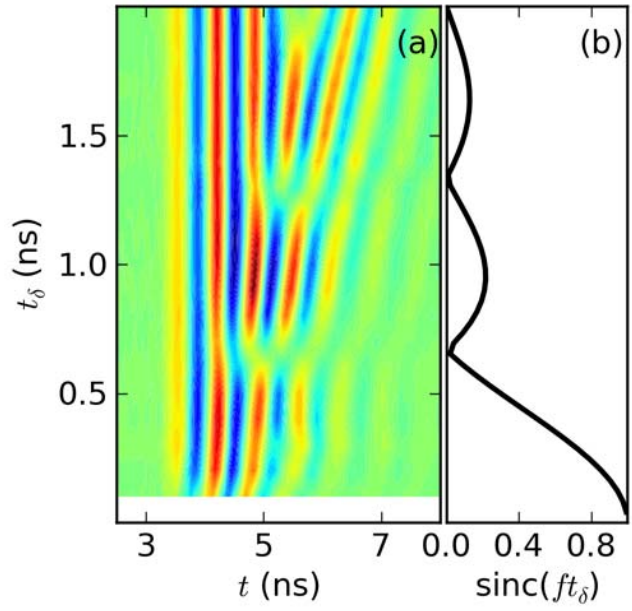

Figure 4.